\documentstyle[aps,epsf]{revtex}

\draft
\newcommand{\beq}{\begin{equation}}
\newcommand{\eeq}{\end{equation}}
\newcommand{\bdis}{\begin{displaymath}}
\newcommand{\edis}{\end{displaymath}}
\newcommand{\bea}{\begin{eqnarray}}
\newcommand{\eea}{\end{eqnarray}}
\newcommand{\barr}{\begin{array}}
\newcommand{\earr}{\end{array}}

\begin{document}

\title{Avalanches in Breakdown and Fracture Processes}

\author{Stefano Zapperi$^1$, Purusattam Ray$^2$,
H. Eugene Stanley$^3$ and Alessandro Vespignani$^4$}

\address{$^1$PMMH-ESPCI, 10 Rue Vauquelin, 75231 Paris cedex 05, France\\
$^2$ The Institute of Mathematical Sciences, C. I. T. Campus, 
  Chennai 600 113, India\\
$^3$Center for Polymer Studies and Department of Physics,
        Boston University, Boston, Massachusetts 02215\\        
        $^4$The Abdus Salam 
International Centre for Theoretical Physics (ICTP) 
P.O. Box 586, 34100 Trieste, Italy}

\date{6 October 1998}

\maketitle

\begin{abstract}

We investigate the breakdown of disordered networks under the action of
an increasing external---mechanical or electrical---force.  We perform a
mean-field analysis and estimate scaling exponents for the approach to
the instability.  By simulating two-dimensional models of electric
breakdown and fracture we observe that the breakdown is preceded by
avalanche events. The avalanches can be described by scaling laws, and
the estimated values of the exponents are consistent with those found in
mean-field theory.  The breakdown point is characterized by a
discontinuity in the macroscopic properties of the material, such as
conductivity or elasticity, indicative of a first order transition. The
scaling laws suggest an analogy with the behavior expected in spinodal
nucleation.

\end{abstract}

\date{\today}
%\begin{multicols}{2}

\section{Introduction}

The breakdown of solids under external forces is a longstanding problem
that has practical and theoretical relevance \cite{breakdown}.  The
first theoretical approach to fracture mechanics dates back to the
twenties with Griffith's theory \cite{griff} which says that cracks grow
or heal, depending on whether the external stress prevail over the
resistance at surface of the crack.  Since the work of Griffith, a great
effort has been devoted to experimentally test the validity of the
theory and to extend it to various crack geometries and boundary
conditions \cite{fract}. The Griffith theory, in spirit, is very similar
to the classical theory of nucleation in first order phase transitions
\cite{nucl}.  In bubble nucleation, a critical droplet will form when
the change in free energy due to the bulk forces exceeds that of the
surface terms.

The analogy between first order transitions and fracture has been
investigated further by numerical model and theoretical calculations.
Several authors suggested that the breakdown point in thermally
activated fracture is analogous to a spinodal point.  Spinodal
nucleation \cite{sn}, contrary to classical nucleation, is characterized
by scaling properties and fractal droplets. Rundle and Klein \cite{rk},
analyzing a Landau-Ginzburg equation for the growth of a single crack,
showed that the system obeys scaling laws expected for spinodal
nucleation.  Selinger et al. \cite{sel1,sel2} have studied the problem
in mean-field theory and by numerical simulation.  They conclude that a
solid under stress is in a metastable state and when the external stress
is raised beyond a critical value, corresponding to the spinodal point,
the system becomes unstable. The nature of the nucleation process in a
stressed solid was studied by Golubovic et al. \cite{gol} using Monte
Carlo simulations. Recently, in the framework of elastic theory, it was
shown \cite{buse} that the point of zero external stress corresponds to
the condensation point in gas-liquid first order transitions.  One of
the ambitious goals of these and other studies is to formulate a
statistical thermodynamics of fracture.

Most of the theoretical studies we have discussed deal with the
situation in which fracture is {\em thermally} activated and {\em
quenched} disorder is irrelevant.  In most realistic situations,
however, the solid is not homogeneous and disorder, in the form of
vacancies or microcracks, strongly affects the nucleation process
\cite{sel2,gol}. For example, cracks may start from different defects and
coalesce \cite{jaeg}, in contrast with the assumptions of Griffith-like
theories.  There are situations, encountered for example in material
testing, in which the system is driven by an increasing external stress
and the time scale of thermal fluctuations is larger than the time scale
induced by the driving. In those cases, the system can be effectively be
considered as being at zero temperature and only quenched disorder is
relevant. This is the situation we investigate in this paper.  It is
also worth to emphasize that the breaking process is in most cases {\em
irreversible}, so opening and closing a crack is not like flipping back
and forth a spin.

The understanding of the breakdown of disordered systems has progressed
to a large extent with the use of large-scale simulations of lattice
models \cite{hr}.  In these models a conductor is represented by a
resistor network and a elastic medium by a spring network or other more
complex discretizations.  The disorder is modeled by random failure
thresholds or by bond dilution. In this way the model retains the long
range nature and the tensorial structure of the interactions, which are
computed solving coupled linear equations.  These models have provided a
good description of geometrical and topological properties of fracture,
leading to the injection in this field of scaling concepts
\cite{hansen2}.  Recently, quasi-static lattice models have been used
also to study dynamical properties of fracture
\cite{th,zrsv,gcalda,zvs}.

The breakdown of a disordered solid is preceded by intense precursors in
the form of avalanches. It has been experimentally observed that the
response (acoustic emission) to an increasing external stress takes
place in bursts distributed over a wide range of scale. Examples are
found in the fracturing of wood \cite{ciliberto}, cellular glass
\cite{strauven} and concrete \cite{ae}, in hydrogen precipitation
\cite{ccc}, in dislocation motion in ice crystals \cite{grasso} and
in volcanic activity \cite{dmp}.  These phenomena are reminiscent of the
Gutenberg-Richter law for earthquakes statistics \cite{gr}.  It is
becoming apparent that avalanche response is rather the rule than the
exception in driven disordered systems. Other examples range from the
motion of domain walls in magnets (the Barkhausen effect) \cite{bark}
and flux lines in superconductors \cite{flux}, frictional sliding
\cite{friction}, to fluid flow in porous media \cite{flow} and the
inflation of degased lungs \cite{lung}.  Therefore, understanding the
general physical mechanisms of avalanche dynamics goes well beyond the
study of breakdown and fracture.

In this paper, we numerically study the random fuse model
\cite{fuse,duxbury,fuse2,fuse3} and a spring network \cite{spring}.  The
investigation of these two models allows us to compare the behavior of a
quasi-static scalar model, with a more complex vectorial dynamical
model.  We analyze the scaling close to the breakdown and we find that
it is consistent with a mean-field analysis. We show that avalanche
behavior near the breakdown in disordered systems is analogous to the
formation of droplets observed close to a spinodal instability in
first-order phase transitions.  The system is driven by a slowly
increasing external force through a complex energy landscape; it is not
allowed to jump over energy barriers by thermal activation.  This
situation should correspond to the experiment reported in
Ref.~\cite{ciliberto}. The intriguing consequence of this analysis is
that the behavior of a disordered driven system at zero temperature is
similar to what is expected from a thermally driven homogeneous system
close to a spinodal point. It is tempting to conclude that quenched
disorder has an effect similar to thermal fluctuations, although a
discussion in terms of metastability and nucleation is not possible in
the first case.  We will briefly discuss these analogies in the magnetic
context.  Finally, it is interesting to remark that spinodal nucleation
and first-order transitions have also been suggested to play an
important role in the physics of frictional sliding and earthquakes
\cite{earth}.  Also for these systems thermal disorder is expected to be
irrelevant with respect to quenched inhomogeneities.

The paper is organized as follows. In section II, we briefly review
spinodal nucleation and we discuss the role of quenched disorder.  In
section III, we present a mean-field analysis of fracture.  Section IV
discusses the simulations for the random fuse model and section V is
devoted to molecular dynamics simulations of a spring network.  In
section VI, we discuss the cluster structure of our models and draw
analogies with percolation and in section VII, we compare the results of
our model with experiments and discuss some open questions.

A short account of a subset of these results has been presented in
Ref.~\cite{zrsv}.

\section{Spinodal Nucleation, Thermal Fluctuations and Quenched
Disorder}

Nucleation near a spinodal appears to be very different than classical
nucleation and the classical theory is expected to fail. Droplets appear
to be fractal objects and the process of nucleation is due to the
coalescence of these droplets, rather than the growth of a single one
\cite{coal}.  The theoretical description of homogeneous spinodal
nucleation is based on the Landau-Ginzburg free energy of a spin system
in presence of an external magnetic field \cite{sn}.  When the
temperature is below the critical value, the free energy has the typical
two-well structure.  In the presence of an external magnetic field, one
of the wells is depressed with respect to the other, which represents
therefore the metastable state. The system must cross a free energy
barrier to relax into the stable phase. When the external field is
increased, this nucleation barrier decreases, eventually vanishing at
the spinodal. Using this formalism, it has been shown that the approach
to the spinodal is characterized by scaling laws, analogous to critical
phenomena.  The magnetization or the order parameter $\phi$ scales with
the external field $H$ as
\beq
\phi-\phi_s \sim (H_s-H) ^{1/2}
\label{eq:op}
\eeq
where $\phi_s$ and $H_s$ are respectively the order parameter
and the field at the spinodal. This law implies a divergence
of the quasi-static susceptibility
\beq
\chi\equiv\frac{d\phi}{dH} \sim (H_s-H) ^{-\gamma};~~~~~~\gamma=1/2
\label{eq:susc}
\eeq
The fluctuations in the order parameter can be related to suitably
defined percolation clusters, whose sizes turn out to be 
power law distributed
with an exponent $\tau=3/2$, in mean-field theory.  For
finite-dimensional short-range models, this mean-field picture is
expected to fail, since the system will nucleate before reaching the
limit of metastability.  On the other hand mean-field behavior is
expected to be valid in presence of long-range interactions, and it has
been numerically verified in Monte Carlo simulations of long range Ising
model \cite{raykl}.  The limit of stability in thermally activated
homogeneous fracture is believed to correspond to a spinodal point.  One
should then be able to observe scaling laws consistent with those found
in spinodal nucleation. To our knowledge, such a scaling behavior has
not yet been observed in numerical simulations.

In this paper we will not deal with thermally-activated fracture but
rather with a disordered driven system.  In this regard, an interesting
analogy can be made with a model recently proposed by Sethna et al.
\cite{sethna,dahm} in the context of magnetic hysteresis.  The model in
question is a random-field Ising model (RFIM), driven at zero
temperature by an increasing uniform magnetic field $H$.  Each spin
$s_i$ takes the sign of the local force
\beq
f_i\equiv-\frac{\delta E}{\delta s_i}=J\sum_js_j+h_i+H,
\eeq
where the sums runs over the neighboring sites, and $h_i$ is the random
field at site $i$, which has a Gaussian distribution with variance $R$.
When the external field is increased, some local forces change sign and
the spins flip along the direction of H in avalanches. For low values of
the strength of the disorder $R<R_c$, there is a critical value of the
field $H_c$ for which the system undergoes a discrete (first-order)
transition involving a finite change in the order parameter (the
magnetization).  

In mean-field theory, the approach to the instability $H_c$, is
characterized by the same scaling laws and exponents of spinodal
nucleation, as reported in Eqs.~(\ref{eq:op})--(\ref{eq:susc}).  Close
to the first-order transition, the avalanche size distribution is
described by a scaling form
\beq
P(m)\sim s^{-\tau}f[m(H_c-H)^\kappa],
\eeq
with $\tau=3/2$ and $\kappa=1$.  It is worth noting that these scaling
exponents coincide with the mean-field exponents for the distribution of
droplets in homogeneous spinodal nucleation.  {}From these studies it
appears that the behavior of thermally activated homogeneous spinodal
nucleation is similar to the approach to the instability in disordered
systems driven at zero temperature.  However, one should bear in mind
that for a given realization of the disorder the dynamics is completely
deterministic in the second case.  Concepts such as metastability and
nucleating droplets are formally not defined in this context.

\section{Mean-Field Theory}

In this section we generalize the analysis of Ref.~\cite{sel1} to derive
a simple mean-field theory for fracture.  The models we will analyze are
defined on a two-dimensional lattice. Each bond of the lattice is
supposed to obey the equations of linear elasticity, until it is
stretched beyond a randomly chosen threshold, after which it breaks.  In
the electric case the equations are scalar, each bond satisfies the Ohm
law, and the currents are computed numerically by solving the Kirchhoff
equations with the appropriate boundary condition.

To illustrate the mean-field theory we will consider for simplicity the
random fuse model.  To every bond $i$ of the lattice we associate a fuse
of unit conductivity $\sigma_i=1$. An external current $I$ or voltage
$V$ is then applied to the system by imposing an external voltage $V$ to
two opposite edges of the lattice.
When the current in the bond exceeds a
randomly distributed threshold $D_i$ the bond becomes an insulator
($\sigma_i=0$).  The voltage drops $(\Delta V)_i$ for each bond are
computed by minimizing the total dissipated energy
\beq
E(\{\sigma\})\equiv\frac{1}{2}\sum_i \sigma_i [(\Delta V)_i^2-D^{2}_i].
\label{en}
\eeq
The dynamics of the model results from a double minimization process.
The voltage drops $(\Delta V)_i$ are obtained by a {\em global}
minimization of the energy at fixed $\sigma_i$, while the $\sigma_i$ are
then chosen to minimize the {\em local} bond energy. The first step is
equivalent to solving the Kirchhoff equations for the network, while the
second step corresponds to breaking the bonds for which the current
overcomes the threshold. The external current is increased slowly until
the lattice is no longer conducting. This means that each time a bond is
broken, the voltage and the currents are recomputed with the new values
of the conductivities.

To derive a mean-field theory, it is useful to recast the dynamics of
the model in terms of the externally applied current $I$. We can rewrite
the energy of Eq.~(\ref{en}), in full generality, as
\beq
E(I,\{\sigma\})=\frac{1}{2}\left(\frac{I^2}{G(\{\sigma\})}
-\sum_i\sigma_i D^{2}_i\right),
\eeq
where $G(\{\sigma\})$ is the total conductivity of the lattice and is a
complicated function of the local conductivities.  We can estimate 
$G(\{\sigma\})$ using the effective medium theory \cite{kirk}, which
in our case gives
\beq
G(\{\sigma\})=2\phi-1,
\label{eq:effmd}
\eeq
where $\phi\equiv\sum_i \sigma_i/L^2$.  We can express the energy as a
sum over ``spins'' interacting with effective random fields $h_i$
\beq
E_{MF}(I,\{\sigma\})=\sum_i\sigma_i h_i=
\frac{1}{2}\sum_i \sigma_i\left(\frac{I^2}{L^2\phi(2\phi-1)}-D_i^2\right).
\eeq
the value of $\phi$ can be computed self-consistently as
\beq
\phi = P(h_i<0)=1-\int_0^{I/[L\sqrt{\phi(2\phi-1)}]}\rho(D)dD, 
\label{mfphi}
\eeq
where $\rho(D)$ is the distribution of failure thresholds.  The solution
of this equation can be expressed in terms of the current per unit
length $f\equiv I/L$.  We can identify $f$ with the external field and
$\phi$ with the order parameter.

{}We can show (see Appendix) that under general conditions $\rho$ in
Eq.~(\ref{mfphi}) has a solution for $f<f_c$ and, close to $f_c$, $\phi$
scales as
\beq
\phi - \phi_c \sim (f_c-f)^{1/2}.
\label{scalphi}
\eeq
The mean-field theory we have presented is very similar to the
fiber-bundle model (FBM) with global load sharing, an exactly solvable
model for fracture which has been studied extensively
\cite{dfbm,hansen1}.  In the FBM an external load $F$, is applied to $N$
parallel fibers, and equally shared among the unbroken ones. This means
that the force on each fiber is
\beq
f_i=F/n
\eeq
where $n=N\phi$ is the number of unbroken fibers. A fiber breaks when
its force exceeds a quenched random threshold $D$.  One can write an
equation for the density of unbroken fibers that has the form of
Eq.~(\ref{mfphi}), with the upper limit of integration replaced by
$F/(N\phi)$.  The FBM can be obtained as a mean-field theory in the case
of site damage, since in this case the effective medium conductivity is
given by $G(\phi)\simeq\phi$.

We can obtain the mean-field avalanche size distribution from the exact
results derived for the FBM \cite{hansen1}
\beq
P(m)\sim m^{-\tau}f(m(f_c-f)^\kappa); ~~~~~\tau=\frac{3}{2},~~\kappa = 1,
\label{eq:pm}
\eeq
where $m$ is the number of bonds that break as function of the current.
Eq.~(\ref{eq:pm}) can also be obtained in the case of 
bond damage using similar arguments.

The average avalanche size $\langle m \rangle$ is proportional to the
``susceptibility'' $d\phi/df$ \cite{chak}, and therefore diverges at the
breakdown as
\beq
\langle m \rangle \sim (f_c-f)^{-\gamma}~~~~~~\gamma=1/2.
\label{susc2}
\eeq
The exponents we have introduced satisfy the scaling relation
$\kappa(2-\tau)=\gamma$, which is consistent with the values reported in
Eq.~(\ref{eq:pm}) and Eq.~(\ref{susc2}).  The mean-field analysis
indicates that the system is undergoing a first order transition since
the order parameter has a discontinuity and the conductivity at $f_c$
has a jump from $G(\phi_c) > 0$ to zero.  The approach to this
transition is characterized by avalanches of increasing size, diverging
at the transition.

A similar behavior with the same scaling exponents is observed in the
mean-field theory of the driven RFIM \cite{sethna,dahm} for small
disorder. In the RFIM, one observes also a second order transition as
the width of the disorder is increased. A similar transition does not
seem to be present in our system, at least not in the mean-field
treatment.  It is also interesting to note that the same scaling laws
describe metastable systems close to a spinodal point.  The quasistatic
susceptibility diverges as in Eq.~(\ref{susc2}) and droplets are
distributed according to Eq.~(\ref{eq:pm}).

\section{Scaling before Breakdown in the Random Fuse Model}

An important issue to address at this point is the validity of
mean-field results in the case of real low-dimensional systems.  It is
known that scaling does not hold close to the first-order transition for
short-ranged RFIM in dimensions $d=2,3$ \cite{sethna,dahm}.  Similarly,
spinodal singularities are observed when interactions are long-range
\cite{raykl}.  Elastic interactions are intrinsically long-range, which
could lead to mean-field behavior even for low dimensions, as we will
next show numerically.

We simulate the random fuse model \cite{fuse} on a tilted square
lattice, with periodic boundary conditions in the transverse direction.
As we discussed in Sec. III, the current in each bond is obtained
solving Kirchhoff equations, which we do numerically using a multigrid
\cite{multigrid} relaxation algorithm with precision $\epsilon = 10^
{-10}$. The distribution of thresholds is chosen to be uniform in the
interval $[1-\Delta,1+\Delta]$~\cite{fuse1}. We impose an external
current $I$ through the lattice and we increase it at an infinitesimal
rate. When a bond fails, we recompute the currents to see if other
failures occur. The process is continued until a path of broken bonds
spans the lattice and no current flows anymore.

\begin{figure}[htb]
\narrowtext
\centerline{
        \epsfxsize=8.0cm
        \epsfbox{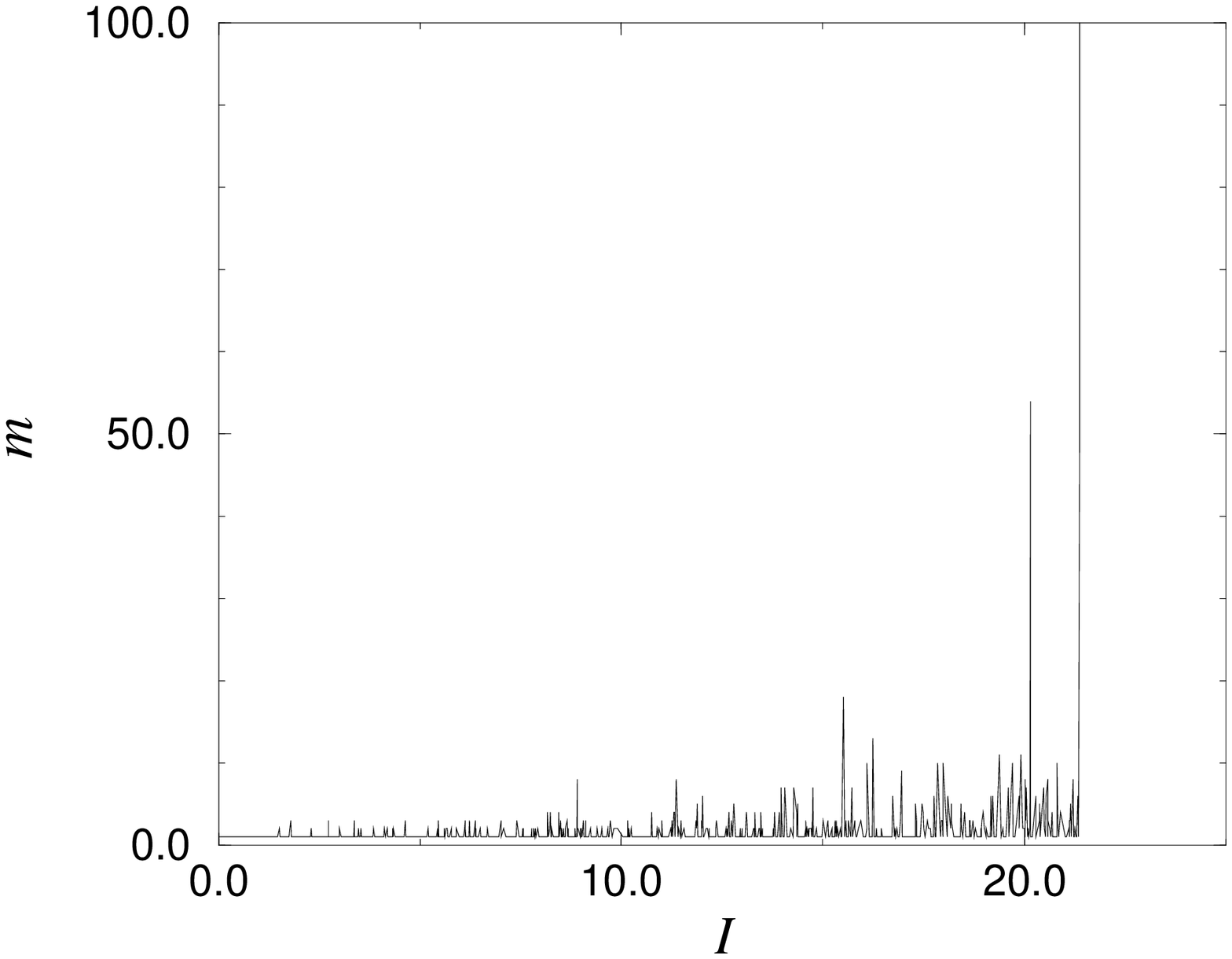}
        \vspace*{0.5cm}
        }
\caption{Avalanches in the fuse model. As the current
is increased the bonds break in avalanches of increasing
size until the final breakdown occurs.}
\label{fig:1}
\end{figure}

\begin{figure}[htb]
\narrowtext
\centerline{
        \epsfxsize=8.0cm
        \epsfbox{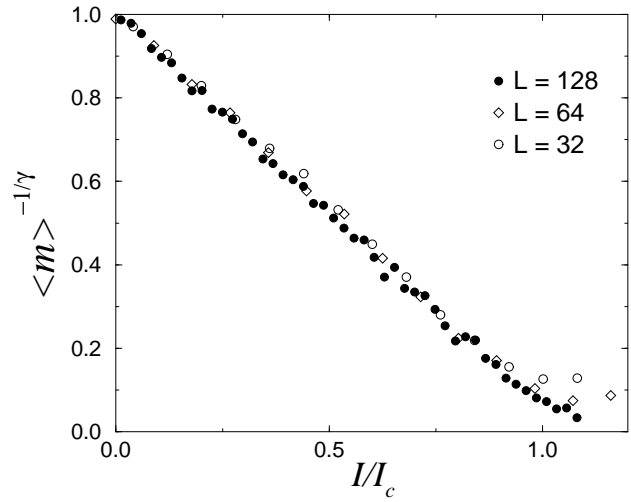}
        \vspace*{0.5cm}
        }
\caption{The average avalanche size in the fuse model scaled 
with the mean-field
exponent ($\gamma=1/2$) as a function of $I/I_c$, for
different values of the system size $L$. The linearity of the plot 
supports the validity of the mean-field calculations.}
\label{fig:3}
\end{figure}

We determine the cluster size distribution $n(s,I)$, which is defined as
the number of clusters formed by $s$ neighboring broken bonds when the
applied current is $I$. The moments ($M_k(I) \equiv \int s^k n(s,I) ds$
is the $k$-th moment) of $n(s,I)$ describe much of the physics
associated with the breakdown process. We determine $n(s,I)$ by
averaging over the various threshold distribution configurations.  The
first moment $M_1(I)$ is the total number of broken bonds due to the
current $I$ and is therefore proportional to $\phi$.  According to our
mean-field picture, the average $\langle m\rangle$ of the quantity $m =
dM_1(I)/dI$ should then diverge close to the breakdown as $(I_c
-I)^{-1/2}$.  In Fig.~\ref{fig:1}, we plot the number of bonds $m$ that
break for a given value of the current, in a particular realization of
the process. We see that the breakdown is highly inhomogeneous with
avalanches of increasing amplitude.  In order to test the scaling, we
plot in Fig.~\ref{fig:3} $\langle m \rangle^{-2}$ as a function of the
reduced current $I/I_c$, where $I_c$ is the average breakdown current,
and we see that the graph is linear.

\begin{figure}[htb]
\narrowtext
\centerline{
        \epsfxsize=8.0cm
        \epsfbox{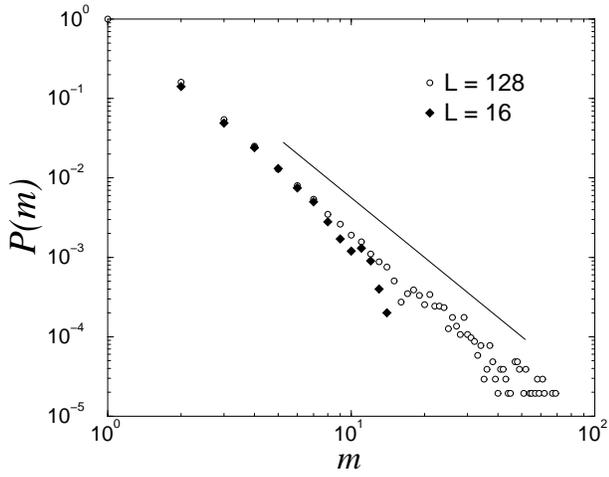}
        \vspace*{0.5cm}
        }
\caption{The avalanche size distributions in the fuse model
for two values of system size plotted in log-log scale. A line with the
mean-field value $\tau^{\prime}=5/2$ of the exponent is plotted for
reference.}
\label{fig:4}
\end{figure}

We also measure the distribution of avalanche sizes, integrating over
all the values of the current. The mean-field analysis predicts that
$P(m,I)\sim m^{-3/2}f(m(I_c-I))$ which yields an exponent
$\tau^{\prime}=5/2$ when the distribution is integrated over the
current. In fact we see that our data are consistent with this exponent
(Fig.~(\ref{fig:4})).  We have also checked that the cut-off of the
distribution increases with the system size.  A similar result
($\tau\simeq 2.7$) for smaller lattice sizes ($L=40$) was previously
reported by Hansen and Hemmer \cite{hansen}, who also pointed out the
similarity with the predictions of the FBM.

\section{Scaling in the fracture of a spring network}

The study of fracture in elastic medium is carried out by molecular
dynamics (MD) simulation on a lattice model with elastic restoring
forces. Our model consists of a $L \times L$ ($L = 20$ and $50$) square
network with central and rotationally invariant bond-bending forces.
The potential energy of the network is \cite{spring}
\begin{equation} 
V = \frac{a}{2}\sum_{\langle ij\rangle}(\delta r_{ij})^2 g_{ij} +    
\frac{b}{2} \sum_{\langle ijk\rangle}(\delta \theta_{ijk})^2
g_{ij}g_{jk}, 
\label{elas}
\end{equation}
where $\delta r_{ij}$ is the change in the length of the spring between
the nearest neighbor sites $<ij>$ from its equilibrium value (taken to
be unity), and $\delta \theta_{ijk}$ is the change in the angle between
the adjacent springs $ij$ and $jk$ from its equilibrium value $\pi/2$
which is taken to ensure the square lattice structure of the lattice at
equilibrium. $g_{ij}= 1$ if the spring $ij$ is present and is 0
otherwise.  $a$ and $b$ are the force constants of the central and the
bond-bending force terms respectively.  In terms of arbitrary length and
time scales $l_0$ and $t_0$, the equations of motion in dimensionless
variables involve the parameters $\lambda_1 = a t_0^2/m$ and $\lambda_2
= b t_0^2/m l_0^2$, where $m$ is the mass associated with the lattice
sites.  We choose $l_0$ to be the lattice spacing (most naturally).  The
ratio $\lambda_2/\lambda_1 = b/a l_0^2$ is then a characteristic of the
system under consideration.  We choose $\lambda_1=1$ and
$\lambda_2=0.1$.  The small value of $\lambda_2$, much less than the
value of $\lambda_1$, allows the fracture to develop without much
deformation of the network. We start with all the springs intact so that
$g_{ij}=1$ for all neighboring $ij$'s and with each spring we associate
a random breaking threshold $D_{ij}$, chosen from a uniform distribution
$D \in [0,2]$.

We impose a constant external force $F$ on the sites of the boundary and
the system is allowed to evolve dynamically using Verlet's algorithm
\cite{spring},
\begin{equation}
\vec{r_i}(t+\Delta t) = 2\vec{r_i}(t) - 
\vec{r_i}(t-\Delta t) + \vec{F_i}(\Delta t)^2.  
\end{equation}
Here, $F_i$ is the force (as determined from Eq. \ref{elas}) and
$\vec{r_i}(t)$ is the position vector of the site $i$ at time $t$.  The
simulation involves discrete time $t$ in steps of $\Delta t$.  If the
simulation runs for $n$ iterations, then the elapsed time
is $t = n \Delta t $ 
while the real time elapsed is $n t_0 \Delta t$.  One way to speed
up the relaxation process would be to choose a large value for $\Delta
t$.  However, there is an upper limit to this value for the iteration
process to remain numerically stable.  This limit is proportional to the
convergence time for the fastest developing components of the stress
distribution, which is generally very small in disordered systems. We
choose $\Delta t = 0.01$.  In addition, we add to the evolution a
small viscous force to damp out excessive oscillations.
In the course of evolution, if any spring
$ij$ stretches beyond its cut-off value $D_{ij}$, the spring snaps
irreversibly and $g_{ij}$ for that spring is set to zero. 
We increase the external force $F$ by small steps and at each step
we compute the number of broken bonds, which constitute an avalanche.
To average over disorder, the simulation is repeated for 50 different
configurations of threshold values $D$.

The fracture in the network takes place in a series of bursts of bond
(spring) breaking processes. In such a burst, bonds break from different
parts of the network and the fracture grows.  We keep track of the
clusters formed by the connected broken bonds which, when spanning the
network, causes its macroscopic breakdown.

\begin{figure}[htb]
\narrowtext
\centerline{
        \epsfxsize=8.0cm
        \epsfbox{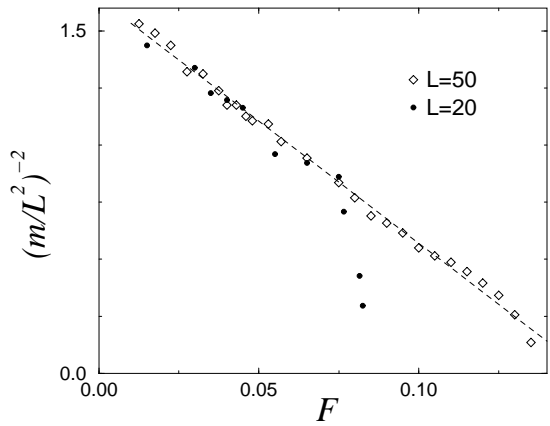}
        \vspace*{0.5cm}
        }
\caption{The average avalanche size in the spring network
scaled with the mean-field
exponent ($\gamma=1/2$) as a function of the applied stress $F$, for
two different values of the system size $L$. The linearity of the plot 
supports the validity of the mean-field calculations.}
\label{ray1}
\end{figure}

\begin{figure}[htb]
\narrowtext
\centerline{
        \epsfxsize=8.0cm
        \epsfbox{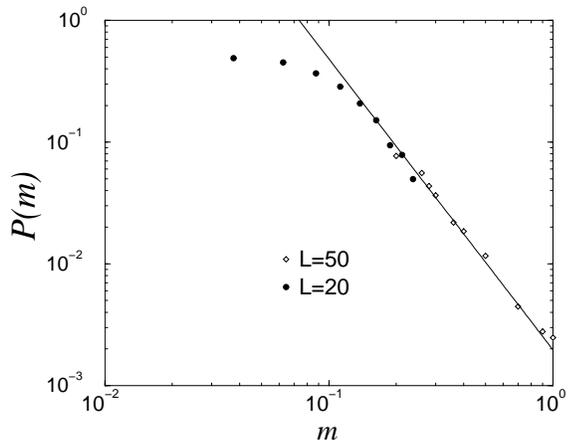}
        \vspace*{0.5cm}
        }
\caption{The avalanche size distributions in the spring network
for systems with two values of system size plotted in log-log scale. A
line with the mean-field value $\tau^{\prime}=5/2$ of the exponent is
plotted for reference.}
\label{ray2}
\end{figure}

We study here the same quantity analyzed in the previous section, namely
the susceptibility and the avalanche distribution.  In Fig.~\ref{ray1},
we plot $\langle m\rangle^{-2}$ as a function of the stress and we find
the linearity as expected.  The critical breaking stress $F_c$ can
readily be found from the points of intersection.  Next, we consider the
distribution $P(m)$ of the value of $m$ integrated over all the values
of $\sigma$ up to $F_c$. We expect the behavior $P(m) \simeq m^{-5/2}$
in the mean-field theory.  Fig.~\ref{ray2} compares our simulation
results with the mean-field prediction, The data presented in the figure
are binned, where for different neighboring $m$-values, the
corresponding $P(m)$-values are combined in one bin, and the result is
plotted as the arithmetic mean of the two extreme $m$-values in the bin.

\section{Cluster Analysis: Nucleation or Percolation?}

The avalanche size defined in the previous sections does not represent
the geometric structure of the cracks, but only counts the number of
bonds breaking at each time step.  A geometrical characterization of the
damage can be obtained studying the cluster size defined counting the
number of connected broken bonds.

\begin{figure}[htb]
\narrowtext
\centerline{
        \epsfxsize=8.0cm
        \epsfbox{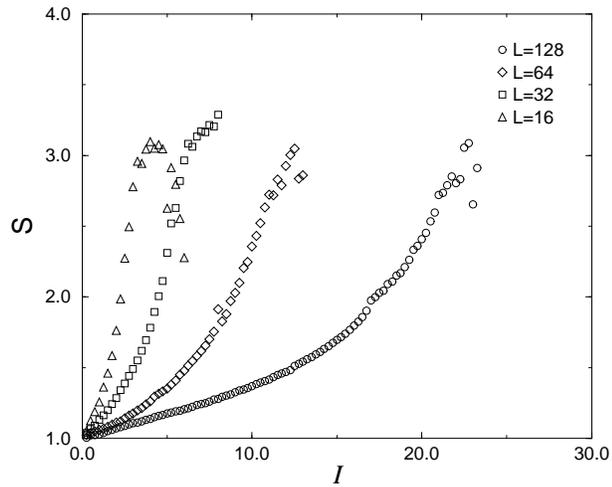}
        \vspace*{0.5cm}
        }
\caption{The average cluster size as a function of the current
for different system sizes. Note that the cluster size
does not diverge.}
\label{fig:5}
\end{figure}

\begin{figure}[htb]
\narrowtext
\centerline{
        \epsfxsize=8.0cm
        \epsfbox{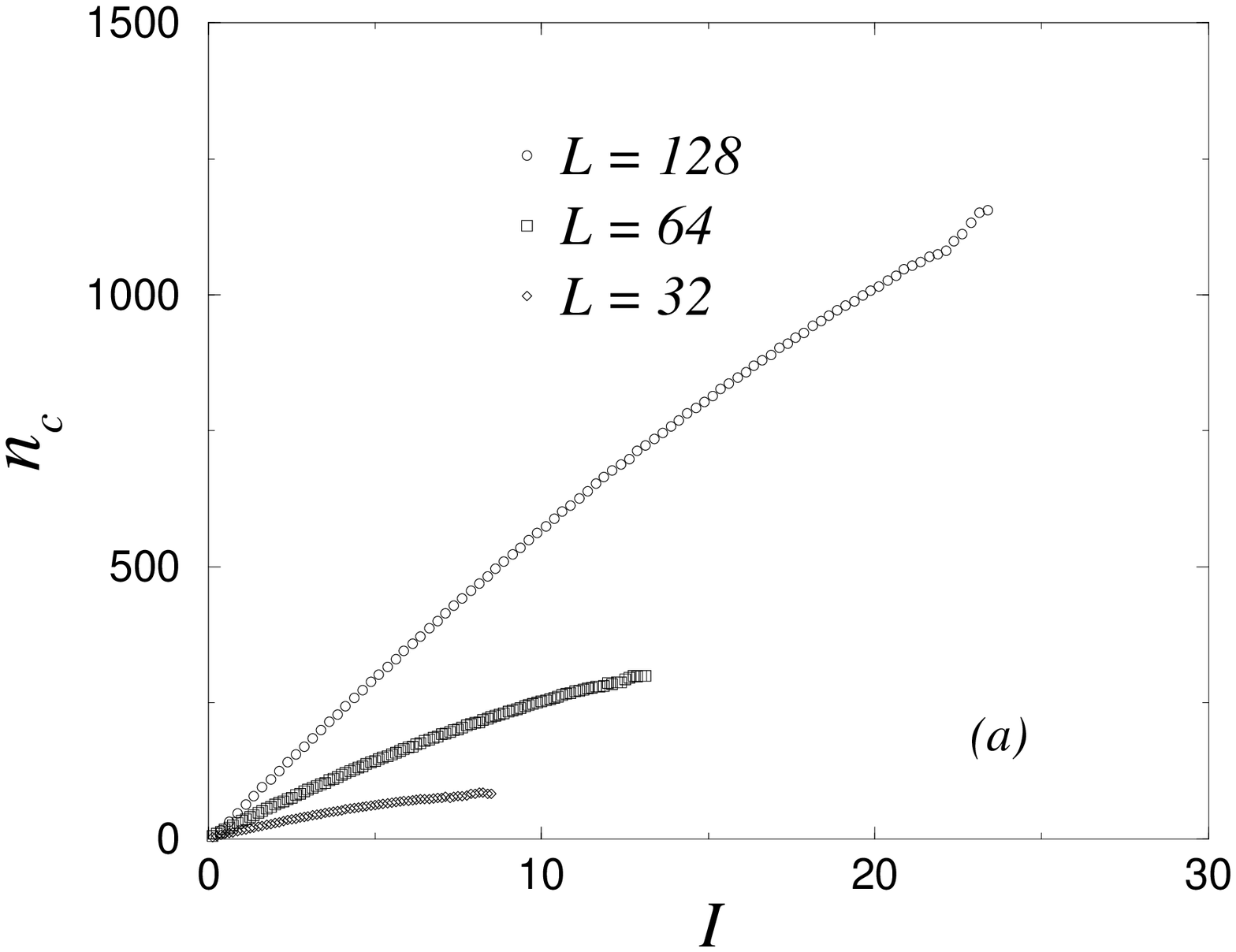}
        \vspace*{0.5cm}
        }
\centerline{
        \epsfxsize=8.0cm
        \epsfbox{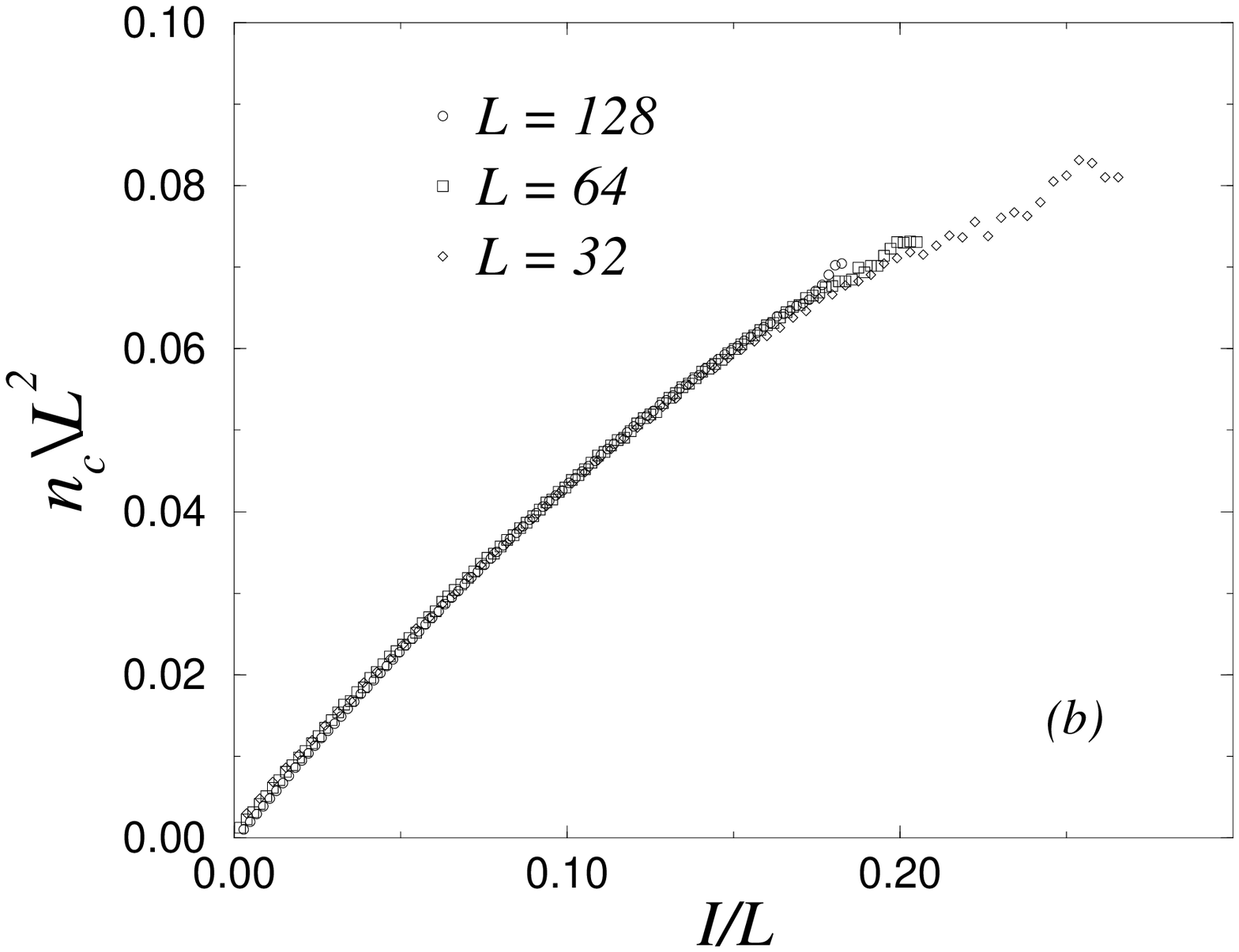}
        \vspace*{0.5cm}
        }
\caption{(a) The number of clusters as a function of the current
in the fuse model for different system sizes.
b) The corresponding scaled plot.}
\label{fig:7}
\end{figure}

\begin{figure}[htb]
\narrowtext
\centerline{
        \epsfxsize=8.0cm
        \epsfbox{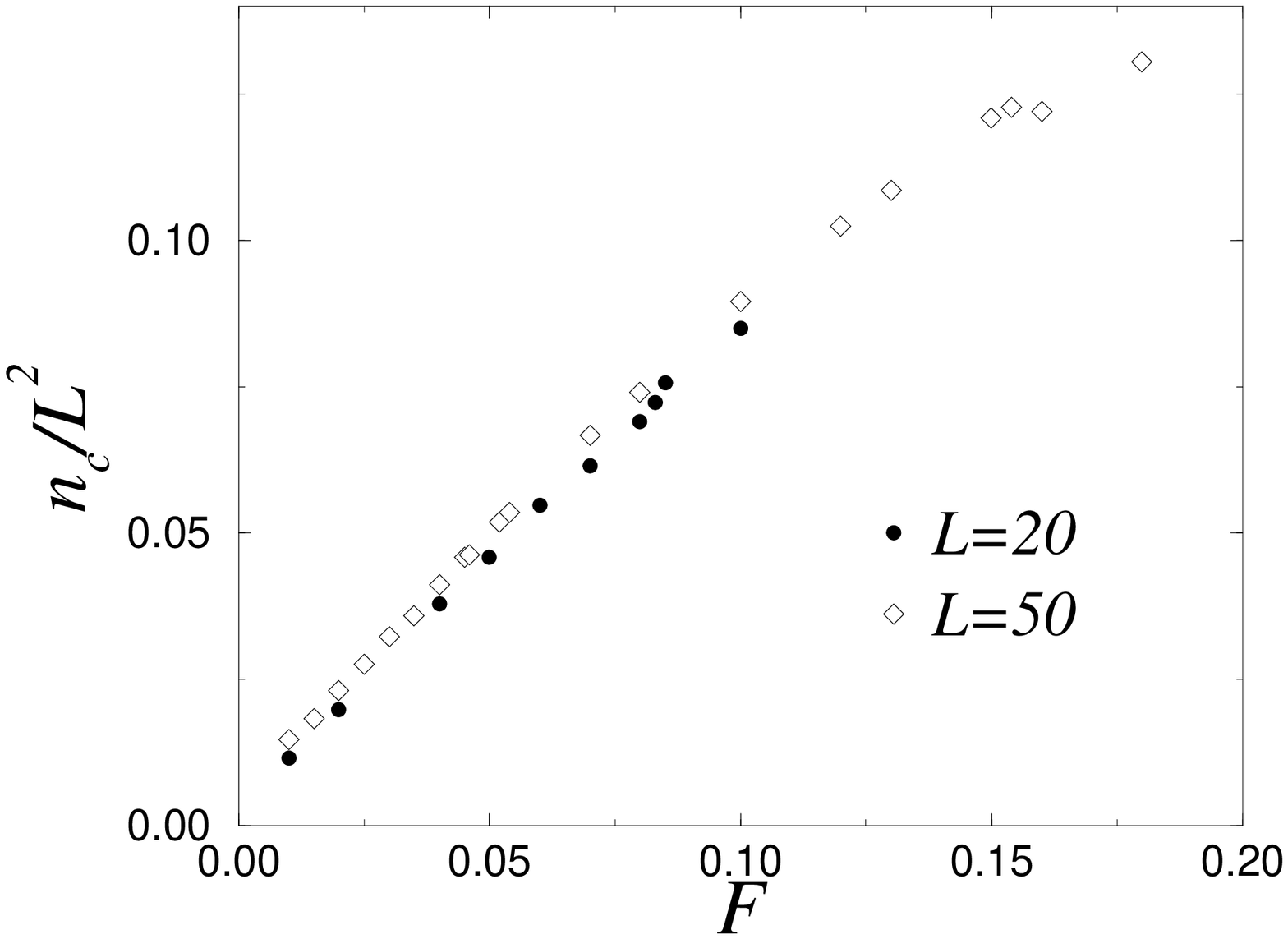}
        \vspace*{0.5cm}
        }
\caption{The scaled number of clusters as a function of the stress
in the spring network for different system sizes.}
\label{fig:a}
\end{figure}

In the fuse model, the average cluster size $S\equiv M_2/M_1$ increases
with $I$. However, by plotting $S$ for different system sizes, we
observe that the cluster size is not diverging (Fig.~\ref{fig:5}).  To
clarify this point, we confirm that $S(I_c)$ does not
show scaling with the lattice size $L$. We find similar results for the
spring network, where the cluster size distribution has an exponential
cut-off that does not change with the lattice size.  We also study the
number of clusters $n_c\equiv M_0$ as a function of the current and for
different system sizes (see Fig.~\ref{fig:7}).  We observe that $n_c$
scales as
\beq
n_c = L^2 g(I/L),
\eeq
which is expected for a first order transition.  We obtain a similar
scaling plot for the spring network (see Fig.~\ref{fig:a}).

\begin{figure}[htb]
\narrowtext
\centerline{
        \epsfxsize=8.0cm
        \epsfbox{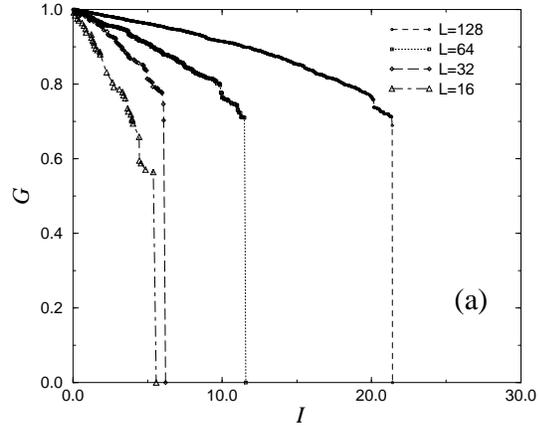}
        \vspace*{0.5cm}
        }
\centerline{
        \epsfxsize=8.0cm
        \epsfbox{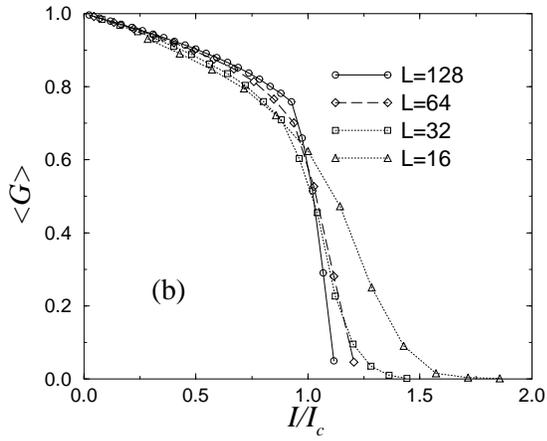}
        \vspace*{0.5cm}
        }
\caption{(a) The conductivity as a function of the current for
a single realization of the disorder, for different system sizes.(b) The
conductivity as function of $I/I_c$ averaged over different realizations
of the disorder. Note that the discrete jump, indicative of a
first-order transition, is smoothed for small system sizes.}
\label{fig:6}
\end{figure}

\begin{figure}[htb]
\narrowtext
\centerline{
        \epsfxsize=8.0cm
        \epsfbox{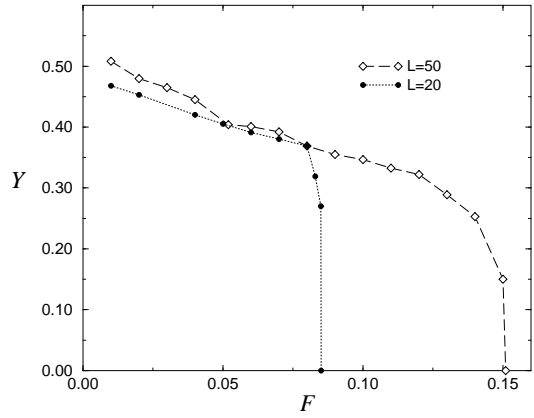}
        \vspace*{0.5cm}
        }
\caption{The elasticity of the spring network as a function of the
  applied stress.}
\label{fig:Y}
\end{figure}

Next we study the behavior of the lattice conductivity in the fuse
model. For a given realization of the disorder the conductivity has a
discrete jump at the breakdown (Fig.~(\ref{fig:6}a).  We also plot the
conductivity averaged over different realizations of the disorder and we
observe a smooth curve with a slope at the breakdown that becomes
sharper as the system size increases (Fig.~\ref{fig:6}b).  In the spring
network, we calculate the lattice elasticity $Y$, which shows similar
behavior as a function of the applied stress (Fig.~\ref{fig:Y}).

Two principal scenarios have been proposed to explain the scaling
behavior of avalanches prior to rupture. The first scenario invokes a
continuous phase transition with a diverging characteristic length
\cite{sor}.  The various cracks inside the lattice should grow until one
of them finally rules over the others, becoming the incipient spanning
cluster. This is precisely what happens in percolation when the
occupation probability $p$ is increased toward the percolation threshold
$p_c$. If this scenario is true for fracture, we would expect the
cluster characteristic size to diverge, contrary to our results.  In the
random fuse network a percolation transition is expected only in the
limit of infinitely wide disorder distributions \cite{roux}, when the
strength of the disorder clearly dominates over the interactions.

The second scenario, in favor of which we presented numerical and
theoretical evidences, describes fracture as a first-order phase
transition close to a spinodal-like instability. The elastic state is
considered to be metastable, as soon as a non-zero stress is applied.
Due to the presence of disorder, the system evolves through a series of
metastable states towards the final instability. This occurs with the
nucleation of cracks growing up to a critical size $s_c$ at which they
{\em coalesce} forming the macroscopic crack. Contrary to percolation,
in this case there is no incipient spanning cluster prior to rupture.

What explains then the scaling in the avalanche statistics and the
susceptibility?  We recall that elastic (or electric) forces are
long-range. When nucleation occurs close to a spinodal
instability---which is well defined only for mean-field or long-range
interactions---one expects a divergent susceptibility \cite{sn}.  This
is not naively related to the fluctuation of a geometrical quantity such
as the crack size, which is not diverging at the spinodal.  In order to
describe geometrically the susceptibility, it is necessary to define the
clusters in a peculiar way, considering each site connected with all the
others within the range of interactions \cite{raykl}.  These
fluctuations are therefore different from those encountered in a
second-order phase transition.

The spinodal point is a quite peculiar critical point which, rigorously
speaking, exists only in mean-field theory, but can be detected when
long-range interactions are present. In this respect, no scaling is
observed in the avalanche distributions when the stress transfer after
breaking is local, such as in the {\em local} load-sharing fiber bundle
model studied in Ref.~\cite{hansen1,hansen}.

\section{Experimental comparison and open problems}

The above results clarify the nature of the breakdown process in the
presence of quenched disorder.  We have found that the breakdown is
preceded by avalanches distributed as power laws. The scaling exponents
are in quantitative agreement with the prediction of mean-field
calculations.  We have discussed that only {\em globally} defined
quantity such as $\langle m(f) \rangle$ and $P(m)$ display scaling,
while locally defined quantities, such as $S$, do not show any singular
behavior. For a second-order phase transition, we would expect local
quantities to show scaling. For instance, in percolation we have $S \sim
(p-p_c)^{-\gamma}$, where $p$ is the concentration of broken bonds.  On
the other hand, first-order transitions usually do not show any
precursor and scaling is not observed.  An exception to this rule is
represented by first-order transitions close to a spinodal point, for
which some global quantities display scaling \cite{raykl}; we will argue
that this case is relevant to the behavior observed before breakdown
\cite{nota_sor}.

The observation that mean-field scaling is present in the fracture of
two different network models suggests that this behavior is rather robust
and does not depend on the fine details of the models, such as the
tensorial structure of the interactions, the dynamics or the boundary
conditions. A more stringent test of our conclusions should 
come from the analysis of experimental data.  In particular,
the experimental setup discussed in
Ref.~\cite{ciliberto} resembles some of the features of our model. An
external pressure is slowly increased until the material (wood
or fiberglass) breaks.
In this process acoustic energy is released in bursts, whose amplitude
shows a net increase as the material approaches the breakdown point
\cite{ciliberto}.  The integrated distribution of burst energies $E$ was
found to follow a power law with an exponent roughly equal to $-2$,
which must be compared to $\tau=5/2$, if one assumes that $m$ in our
paper is proportional to $E$ in Ref.~\cite{ciliberto}.  We see that
there is a discrepancy in the results, although this may be due to
the statistics. We have also tried to analyze the scaling of
the average energy released at pressure $P$, but we could not
obtain a firm conclusion due to the large statistical uncertainties.
Interesting results have been recently obtained 
in three dimensional simulations of fuse networks \cite{alava}.

In mean-field theory, driven disordered systems behave similarly to
their homogeneous, thermally driven, counterparts, if we compare the
scaling of avalanches with that of the droplets.  This applies to the
RFIM \cite{sethna,dahm}, which shows features similar to those of
spinodal nucleation \cite{sn}, and to the fracture models we have
studied. However, one should be careful not to interpret these analogies
too strictly, since in driven disordered systems the notions of
metastability, spinodal point and nucleation are not well defined.  In
particular, the identification of $\phi$ and $f$ with the order and
control parameters is justified only in MF theory.  In two dimensions,
simple homogeneous scaling fails in the presence of disorder. The
breakdown current $I_c$ has a logarithmic size dependence
\cite{duxbury},
\beq
I_c \sim \frac{L}{\log L},
\eeq
which cannot be interpreted in MF theory. A similar dependence is also
present in the fraction of broken bond before breakdown $\phi_c$. 
While for finite systems the picture we have presented is completely
consistent, it is not obvious how to perform the $L\to\infty$ limit. In
order to obtain intensive parameters in this limit one should rescale
$f$ and $\phi$ by an appropriate logarithmic factor. This can be done
implicitly analyzing the data in terms of $f/f_c$, as in
Ref.~\cite{ciliberto}.  Finally, we note that in most cases the final
breakdown starts from existing defects in the material. In our
simulations, we restricted our attention to the case in which
these defects were smaller than the discretization unit. 

In conclusion, we have shown that two different models of breakdown and
fracture share the same mean-field scaling exponents approaching the
rupture point.  The behavior observed in these models is analogous to
spinodal nucleation in thermally driven homogeneous systems. At the
breakdown point, the macroscopic quantities (elasticity, conductivity)
are discontinuous and the characteristic crack size $S$ stays finite in
the large $L$ limit. In addition, the statistics of global quantities
(i.e. the number of broken bonds) display clear mean-field scaling in
analogy with spinodal nucleation. The direct applications of these
ideas to experiments still remains an open question.

\section*{Acknowledgments}
The Center for Polymer Studies is supported by the NSF.  
S. Z. acknowledges financial support from EC TMR Research Network
contract ERBFMRXCT960062. A. V. and S. Z. acknowledge partial
support from from the EC Network contract ERBFMRXCT980183.
We thank G. Caldarelli, 
P. Cizeau, S. Ciliberto, A. Guarino, A. Hansen, H. J. Herrmann, 
W. Klein, A. Petri, S. Roux and F. Sciortino for interesting
discussions and useful remarks. We are particularly grateful to S.
Ciliberto and A. Guarino who kindly provided us with the data of their
experiments.

\section*{Appendix}
We derive here the scaling law for the approach to the instability in
mean-field theory.  We define $f\equiv I/L$ and
$h(\phi)\equiv \sqrt{\phi (2\phi-1)}$ and we rewrite Eq.~(\ref{mfphi}) as
\beq
\phi =1-\int_0^{f/h(\phi)}\rho(D)dD. 
\label{mfphi2}
\eeq
By taking the derivative over $f$ on both sides of the equation,
we obtain for the susceptibility
\beq
\frac{d\phi}{df}=-\frac{\rho(f/h(\phi))}
{(1-\rho(f/h(\phi))f h^{\prime}/h(\phi)^{2})}.
\eeq
When the denominator is equal to zero, the system reaches the
instability and the susceptibility diverges, which defines the critical
values $\phi_c$ and $f_c$
\beq
(1-\rho(f_c/h(\phi_c))f_c 
h^\prime(\phi_c)/h(\phi_c)^{2})=0.
\label{crit}
\eeq
The Taylor expansion of Eq.~(\ref{mfphi2}) around $(\phi_c,f_c)$
yields
\[
\delta\phi\equiv\phi-\phi_c=\rho(f_c/h(\phi_c))
(f_c h^\prime(\phi_c)\delta\phi/h(\phi_c)^2\]
\beq
-\delta f/h(\phi_c))+ 
A\delta\phi^2+B\delta f\delta\phi,
\eeq
where $\delta f\equiv(f-f_c)$ and $A,B$ are two
constants. The term proportional to $\delta\phi$ vanishes because of
Eq.~(\ref{crit}), leaving an equation of the form
\beq
\delta f \sim\delta\phi^2
\eeq
which is the scaling relation reported in Eq.~(\ref{scalphi}).  This
relation thus holds for any analytic normalizable distribution function,
but it is also true in the case of a uniform distribution, as can be
easily shown by a direct calculation.

\newpage



%\end{multicols}

\begin{thebibliography}{99}
\bibitem{fuse1}
  We made the choice $\Delta=1$ in order to avoid the ``ductile-brittle''
  crossover at a finite value of the lattice size 
  (see Ref.~\protect\cite{fuse3}).
  Other broad analytic distributions give rise to similar results.
\end{thebibliography}
\end{document}